\documentclass[a4paper]{article}
\usepackage{textcomp}
\usepackage[noadjust]{cite}

\usepackage{authblk}
\usepackage{graphicx}

\usepackage{amsmath}

\usepackage{bm}

\usepackage{array}

\title{Efficient machine learning for motion sensing for lighting applications}

\author{Fetze~Pijlman}
\affil{Signify Research, \\ Technical University Eindhoven}

\begin{document}

\maketitle

\begin{abstract}
The use of machine learning for building a classifier in signal processing for motion sensing presents unique challenges. This paper proposes a novel method that effectively addresses the combination of skewed data sets and optimization requirements. By utilizing a customized loss function and a product of probability models, our approach achieves a fully automated and efficient machine learning process. Additionally, our resulting probability models offer reduced complexity, making them ideal for embedded applications. Our method offers a promising solution for motion sensing applications that require accurate and efficient classification.
\end{abstract}

\section{Introduction}

\subsection{Motion sensing for lighting applications}

Motion sensing is a popular method for controlling lights in the environment. A popular control scheme contains a first ingredient being that when motion is
observed while the lamp being off, the lamp will be switched on for a predefined length of time (also known as hold time which is tracked by a counter). 
Typical values for hold time are in the range of 5 to 20 minutes. A second
ingredient is that when motion is being observed while the lamp being on, the counter that tracks the amount of time that has passed is being reset. The requirements for the
first ingredient are demanding. For a person approaching the response time should be quick and false positives must be avoided. A typical response time is 0.6 seconds and a
typical false positive rate is 1 per 1000 hours. A low false positive rate prevents a large office building to appear as a disco to an outsider late in the evening. The
requirement for the second ingredient is often substantially different. The motion that needs to be sensed is often smaller in signal (turning a page in a book often leads to
a smaller signal than a person walking in), response time can be slower as a reset of the counter cannot be directly observed, and the false positive rate can be higher since
false positives only lead to an elongation of the time that the light is on. The detectability of a nearby arm motion should typically be above 25\%~\cite{nema} (one is
allowed to miss some motions without leading to drastic consequences) and the false positive rate should typically be below 1 per 20 hours.

\begin{figure}
\centering
\begin{tabular}{ccc}
\includegraphics[width=0.3\textwidth]{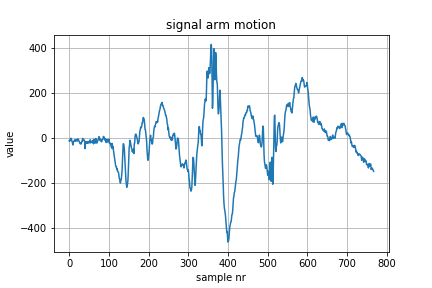} \
\includegraphics[width=0.3\textwidth]{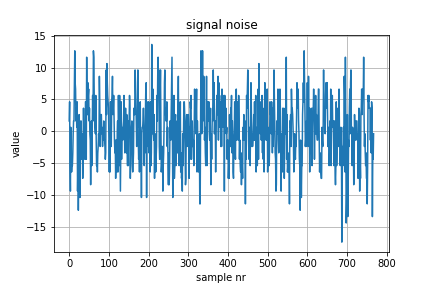} \
\includegraphics[width=0.3\textwidth]{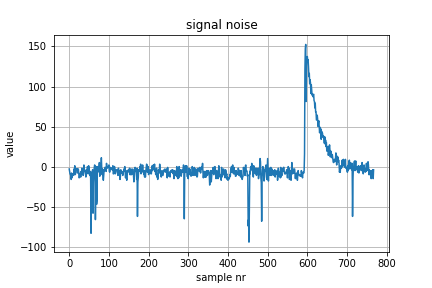}
\end{tabular}
\caption{Some signals related to motion sensing. The left and center graph are typical motion and noise signals, the right graph is an example of a spurious noise
signal.\label{signals}}
\end{figure}

When developing motion sensors one usually records a set of motion events and a substantial amount of hours of so-called noise. The motion events are typically recorded at
various angles and distances from the motion sensor. A data set containing a few thousands of these recorded events is very typical for developing a motion sensor. The
collection of noise data is different. Where the collection of motion data can be done by following a protocol, the collection of noise data is often done by
long measurements. The signals that are collected in these measurements are most often small in size but the appearance of occasional spurious noise signals (e.g. due to radio
interference) is common. Some example signals are presented in Fig.~\ref{signals}.
The recording and handling of occasional spurious noise signals makes the development of a signal processing algorithm challenging.

\subsection{Challenges for using machine learning}
A typical signal processing algorithm is fed by a time series of regularly sampled samples on which the algorithm needs to respond. Algorithms such as recurrent neural
networks and Bayesian models
are of interest although being of the infinite response type (classification depends on full history). Another option is to group samples and to classify the group, an approach that 
has simularities with FIR filters. The latter we will study in this paper.

The collected data for motion and noise are quite different in size. The amount of motion data is typically a few thousand events multiplied by a typical time for motions
(order seconds), making the total amount of motion data to be less than an hour (typically). This is in contrast with the amount of noise data which can easily exceed ten
thousand hours. If one decides to chop the recorded data into smaller parts, where each part either contains a motion event or pure noise, then the resulting data set is
strongly skewed.

There is a more important aspect being
that most of the noise data does not seem to contain valuable information. As pointed out earlier, for the majority of the noise data the signal is small in value. The only
reason why we have a lot of noise data is to capture a small set of spurious signals. This means that training a machine learning model on all the data is inefficient.

\begin{figure}
\centering
\includegraphics[width=0.35\textwidth]{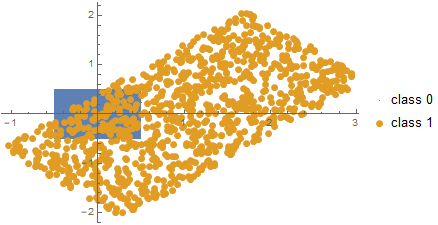}
\caption{Samples taken from two hypothetical distributions. For class 0 we have 100.000 samples and for class 1 we have 1000 samples.
\label{distributions}}
\end{figure}

\begin{figure}
\centering
\includegraphics[width=0.4\textwidth]{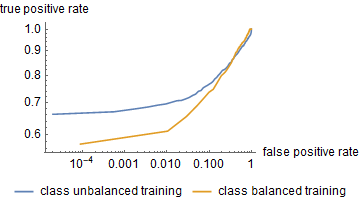}
\caption{Receiver operating curves for a linear logistic model ($P(x, y) = S(a x + by +c)$ where $S$ is a sigmoid function, when using different class weights during training and the binary cross entropy
as a loss function (for an introduction into machine learning please see Ref.~\cite{lecnotes}).
\label{roc}}
\end{figure}

Next to having an imbalanced data set we also need to deal with imbalanced requirements. E.g., when chopping the signal into five second time series the true positive
requirement may be 25\% and the false positive requirement may be 0.007\% (~1/20hrs). One popular method for tuning a probability/machine model to a set of requirements is
to train a model using a generic loss function after which a threshold is tuned: a probability above the threshold may be called motion and below the threshold may be called
noise. 
However, as is shown in a simple example, Fig.~\ref{distributions} and Fig.~\ref{roc}, an optimal classifier requires more than just a tuning of thresholds; choosing different weights
for the class samples leads to different false and true positive rates. As training with different weights is rather time consuming, we strive for a process in which the tuning step can be avoided when the requirements are known upfront.
It is the object of this article to explore machine learning methods to efficiently train algorithms with imbalanced data sets and imbalanced requirements, without the need
of any manual tuning.

\section{Efficient machine learning on requirements}

Optimizing directly on the requirements means that a loss function needs to be constructed in which requirements are compared against actual performance. When a certain
requirement is unmet we will add a penalty with a value that is determined by the amount of mismatch. One problem with this approach is that the mismatch is a priori not continuous as the
number of samples is discrete. In order to use methods like gradient descent, we will make a continous version of the requirements by introducing parameters $\lambda$. As we will
see later these parameters help us to avoid local minima and increase robustness of the solution.

For obtaining a more efficient training method we can limit ourselves in how to filter out the easy noise samples. Our approach will be to train a product of classification models
which together are responsible for
a final classification. The idea is that a first model differentiates between "easy noise" vs "motion and complex noise" while a second model can differentiate between
"complex noise" vs "motion".
For a fully automated approach the classification of "easy noise" vs "motion and complex noise" should be an outcome of the training procedure. This approach
can be extended by adding more models.
In order to make the efficiency gain, the first model may be trained on a subset of the data while the second model may be trained only on positive outcomes on the first
model on all data.

\subsection{Loss function construction}

\begin{figure}
\centering
\begin{tabular}{cc}
\includegraphics[width=0.35\textwidth]{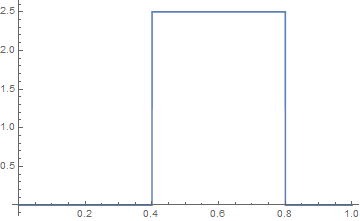} &
\end{tabular}
\caption{An example of the defined spline $s$ being $s(y, 0.6, 0.2)$.\label{spline}}
\end{figure}

For obtaining a continuous loss function the number of false and true positives need to be approximated by a continuous function. In order to do so, the predicted probability is effectively widened, defining
 the fractional positive of a single sample as
\begin{equation}
{\rm fp} = \int_{1/2}^{\infty} {\rm d}y\ s(y, y_{\rm pred}, \lambda),
\end{equation}
where 
\begin{equation}
s(y, y_{\rm pred}, \lambda) = \frac{\theta(y-(y_{\rm pred}-\lambda))\ \theta((y_{\rm pred}+\lambda)-y)}{2\lambda}
\end{equation}
where $\theta$ function represents the Heaviside step function. An example of os $s(y, y_{\rm pred}, \lambda)$ is presented in Fig.~\ref{spline}.

The definition above has two important properties. The first property is that in the limit of $\lambda \rightarrow 0$ the fractional positive becomes the regular number of positive (either 0 or 1). The second property is that the fractional positive equals 0
when $y_{\rm pred} < 1/2-\lambda$. When defining a loss function in terms of fractional positives (as will be done below), this means that the gradient of the loss function disappears for all
samples having a fractional positive equal to 0.

For a complete dataset we proceed by defining fractional false positives as
\begin{equation}
{\rm FFP}(w) = \sum_{i \in \rm class\ 0} \int_{1/2}^\infty {\rm d}y\ s(y, p(x_i,w), \lambda_0),
\end{equation}
and we define fractional true positives as
\begin{equation}
{\rm FTP}(w) = \sum_{i \in \rm class\ 1} \int_{1/2}^\infty {\rm d}y\ s(y, p(x_i,w), \lambda_1).
\end{equation}
Using these definitions the loss function $E$ is as
\begin{eqnarray}
E(w) &=& \sqrt{ E^2_{\rm FP}(w) +  E^2_{\rm TP}(w) },\\
E_{\rm FP}(w) &=& \left( \frac{{\rm FFP}(w)}{{\rm FP}_{\rm req}} - 1 \right)\ \theta \left( {\rm FFP}(w) - {\rm FP}_{\rm req}\right),\\
E_{\rm TP}(w) &=& \left( \frac{{\rm TP}_{\rm req}}{{\rm FTP}(w)} - 1 \right)\ \theta \left( {\rm TP}_{\rm req}-{\rm FTP}(w) \right).
\end{eqnarray}

\subsection{Model and Training Strategy}

The classification model is written as 
\begin{equation}
p(x,w) = S(w_{2a} L(p_1(x, w_1)) + w_{2b})\ p_2(x,w_{2r}),
\end{equation}
where $S$ is the Sigmoid function and $L$ is the logit function.

In the first phase the model $p_1(x,w_1)$ can be trained in isolation using a subset of the data and a class balanced binary cross entropy function. This should give a reasonable classifier that distinguishes between 
easy noise and more complex noise/motion. The choice for a class balanced loss function is motivated by the observation
that a classifier can be constructed from subclassifiers on uncorrelated signals $x_1$, $x_2$ and $x_2$ as follows
\begin{equation}
p(y|x_1,x_2,x_3) = \frac{p(y|x_1)}{p(y)}\ \frac{p(y|x_2)}{p(y)}\ p(y|x_3).
\end{equation}
The equation shows that every additional subclassifier is balanced by the class abundances.

In the second phase 
the weights $w_1$ are frozen, hence the values $L(p_1(x, w_1))$ are frozen as well. This means that for every $w_2$ one can quickly compute
$S(w_{2a} L(p_1(x, w_1)) + w_{2b})$ for the training noise samples. All noise samples having an $S(w_{2a} L(p_1(x, w_1)) + w_{2b})$ below $1/2-\lambda_0$ will not contribute to the error function as $p_2(x,w_{2r})<1$. This means that the training of $p(x,w)$ by varying $w_2$ can be done very efficiently. The filtering effect of easy noise samples is altered during the training
by varying $w_{2a}$ and $w_{2b}$.

\section{Implementation Example and Results}

\subsection{Data}
We use the following data set:
\begin{itemize}
\item motion: 869 $\times 3.2$ second time series (which have been sampled at 240 Hz),
\item noise: 8409679 $\times 3.2$ second time series (also at 240 Hz).
\end{itemize}
The data is split in a train ($2/3$) and a validation ($1/3$) part.

\subsection{Requirements}

The target requirements are chosen to be
\begin{itemize}
\item True positive fraction should be higher than 50\%.
\item False positive fraction should be lower than 0.0044\% (less than 1 per 40 hours).
\end{itemize}
For the training samples the required minimum number of true positives becomes 291 out of 582 motion samples and the maximum required false positives
becomes 125 out of 5624484 noise samples.

\subsection{Features}

The incoming samples are grouped.
Each group of 768 sampled values is translated in a set of features. This translation is done as follows:
\begin{enumerate}
\item Subtract mean from 768 samples.
\item Using a sliding Hanning window of 128 points with step size of 32, 21 real FFTs are computed per group
\item This leads to 21 amplitude ($A$) spectra of which each amplitude spectrum has 64 bins with positive values (zeroth bin can be dropped).
\item Prior to training, all features are normalized to the mean power of the mean power spectrum for a random subset of the data.
\end{enumerate}
In a future study one may want to take the complex part of the FFT into account which is useful for detecting discontinuities.

\subsection{Model}

\subsubsection{$p_1(x, w_1)$}

The first model is a rather simple filter. For distinguishing between "easy noise" and "complex noise and motion" a simple energy filter will be used.
\begin{eqnarray}{rCl}
\left< E \right> &=& \frac{\sum_{i=1}^21 \sum_{j=1}^{64} A_{ij}^2}{21*64}\\
p_1(x, w_1) &=& S(w_{11} \left<E\right> + w_{12}).
\end{eqnarray}

\subsubsection{$p_2(x, w_2)$}

\begin{figure}
\centering
\includegraphics[width=0.12\textwidth]{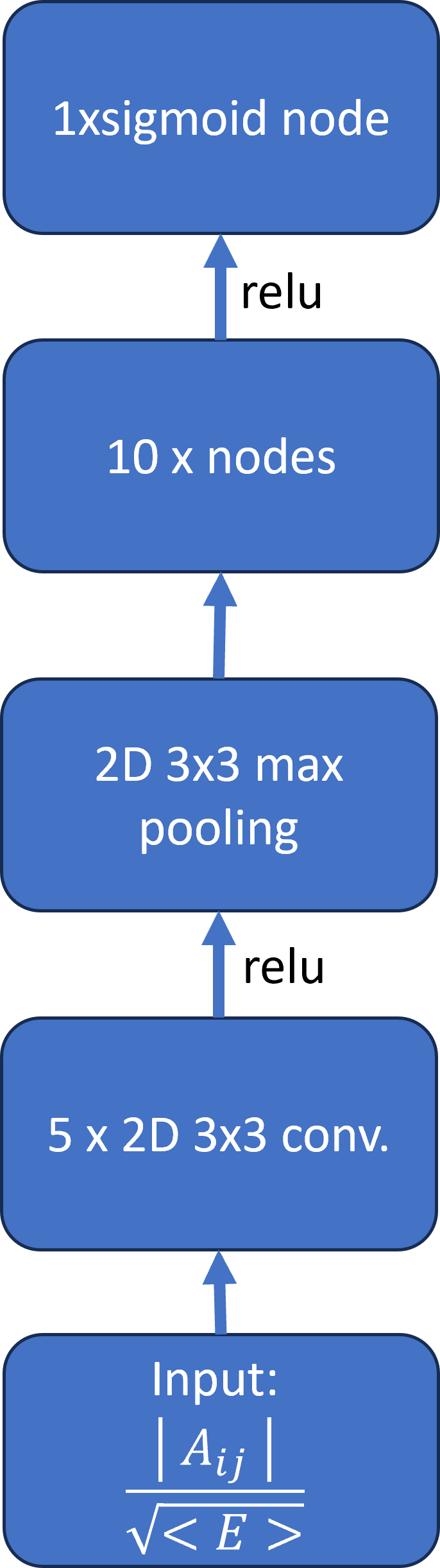}
\caption{Neural network architecture representation of $p_2$.\label{network}}
\end{figure}

The second model is a bit more complex as it needs to distinguish between complex noise and motion. For illustrating the benefits of this method, an adapted simple neural
network was chosen that can also be used for recognizing hard written digits~\cite{nnexample}. The network is built up as follows (see also Fig.~\ref{network}):
\begin{enumerate}
\item 2D convolution layer with 5 types of 3x3 filters with unit norm as kernel constraint and bias constraint between 0 and 1. Activation function is rectified linear unit.
\item 2D max pooling layer with pool size 3x3.
\item Flattening layer
\item Layer with 10 nodes with weights and bias constrained between 0 and 1 and activation function being rectified linear unit.
\item Layer with 1 nodes with weights and bias constrained between 0 and 1 and activation function being a sigmoid function.
\end{enumerate}

\subsection{Training results}

All calculations were done by using keras~\cite{keras} and tensorflow~\cite{tensorflow}. In Table~\ref{benchmarktraining} results from a standard training procedure are given.
Unbalanced training gave no sensible results which can be explained by the fact that in each batch of 32768 samples one has about 50 motion samples. This ratio
apparently insufficient for a succesful training of the model. When equally balancing the classes the training procedure does converge (after 2 epochs).  However, the performance is still
far away from the stated requirements. As shown in Fig.~\ref{roc} the chosen class weights can influence the performance in general. For this data set unbalanced training is apparently not feasible.
Class balanced training is feasible but no satisfactory results were obtained.

\begin{table}
\centering
\resizebox{\columnwidth}{!}{%
\begin{tabular}{l|r|r}
&no class balancing & class balancing\\
\hline
loss 						& $0.0086$ ($0.0086$) 	& $0.47$ ($0.47$)\\
perc. motion (threshold $0.0081$) 	& $0$ ($0$) 			&  \\
perc. noise (threshold $0.0081$) 	& $0.0019$ ($0.0022$)   &  \\
perc. motion (threshold $0.99999985$) &  				& $0$ ($0$) \\
perc. noise (threshold $0.99999985$) 	&  				& $0$ ($0$)\\
perc. motion (threshold $0.65$) 		&  				& $49$  ($50$) \\
perc. noise (threshold $0.65$) 		& 				& $0.83$  ($0.83$)
\end{tabular}
}
\caption{Results on test set and (training set).
Learning rate 0.1, batch size 32768 (no randomization), loss function binary cross entropy.~\label{benchmarktraining}.}
\end{table}

\begin{table}
\centering
\resizebox{\columnwidth}{!}{%
\begin{tabular}{l|r|r|r|r|r|r|r}
nr & model & trained & $\lambda_0$ & $\lambda_1$ &  loss &  perc. TP &  perc. FP \\
\hline
1 & $p_1(x, w_1)$ & $w_1$ & NA & NA 			& $0.43^*$ ($0.43^*$) 	& $60$ ($60$) 	& $1.6$ ($1.6$)\\
2 & $p(x,w)$ & $w_2$ & $0.49$ & $0.49$ 		& $0.23$ ($0.22$) 	& $40$ ($41$)	& $0.0016$ ($0.0016$)\\
3 & $p(x,w)$ & $w_2$ & $0.245$ & $0.49$		& $0.14$ ($0.12$) 	& $42$ ($45$)	& $0.0017$ ($0.0021$)\\
4 & $p(x,w)$ & $w_2$ & $0.1225$ & $0.245$		& $0.0024$ ($0.$) 	&  $49$ ($50$)	& $0.0011$ ($0.0014$)
\end{tabular}%
}
\caption{Results at various stages for the test set and (training set). The loss value $(*)$ of the first stage is the unbalanced binary cross entropy~\label{tabletraining}.}
\end{table}

In the following results from the training procedure in conjuction with the special loss function will be discussed.
The purpose of model $p_1$ is to separate the "easy noise" from the "complex noise/motion". To this end it is sufficient to train the model on a random subset of the noise data
as this should be dominated by "easy noise".
A training data set for model $p_1$ was constructed by taking all motion samples from the training data set supplemented with a random subset of the noise data making a complete set of 65536 samples. 
The weights $w_1$ were optimized using a class balanced binary cross entropy function. The results can be found in Table~\ref{tabletraining}.

For training weights $w_2$ the constructed loss function was used combined with the complete training data set of 562k samples. 
At each iteration samples were selected that had a nonzero contribution to the loss function making
the training efficient. The number of samples varied during the training but when a maximum of 65536 samples was reached the training was stopped.
The loss function is parametrized through $\lambda_0$ and $\lambda_1$. The start values for $\lambda$ parameters were taken to be large as this
maximally reduced local variations in the loss function and thereby reduces the possibility of entering a local minimum. Upon convergence
or reaching the maximum number of training samples, the $\lambda$ parameters were reduced by a factor of $2$.

In Table~\ref{tabletraining} results at the various stages are given. Overfitting did not occur for any of the training strategies. Moreover, the last stage does show that it is possible to meet the
desired requirements that are needed for the use case to detection motion when the lights are already on. More work is needed to see if requirements can be met for motion detection when the lights are switched off.

\section{Conclusion and discussion}

This paper presents an approach to address the challenges of using machine learning with signal processing algorithms for motion sensing. Imbalanced data sets and requirements can make standard approaches inefficient and suboptimal. Our proposed approach achieves better performance and substantially reduces training time compared to straightforwardly minimizing the binary cross entropy. We tested this approach on microwave motion sensing data, using no more than 65k samples out of 600k in training, leading to a significant reduction in training time by a factor of 100. Additionally, the trained model required no tuning of thresholds.

Our approach uses multiple models combined with a loss function that excludes well-classified samples from the gradient calculation, making it more focused on classifying rather than probability fitting. This allows us to use less complicated models that describe only a fraction of the sample space, reducing the risk of overfitting and making them more easily deployable on simple microprocessors.

To optimize directly on requirements and maximize the separation of probability densities of different classes, we used controlled smoothed distributions as input to the loss function. This concept is similar to support vector machine methods~\cite{svm}, in which the goal is to maximize the gap between samples of different classes. By using this approach, we achieve maximum protection against overfitting in the absence of more information.

\section{Acknowledgements}
EdgeAI “Edge AI Technologies for Optimised Performance Embedded Processing” project has received funding from Key Digital Technologies Joint Undertaking (KDT JU) under grant agreement No 101097300. The KDT JU receives support from the European Union’s Horizon Europe research and innovation program and Austria, Belgium, France, Greece, Italy, Latvia, Luxembourg, Netherlands, Norway.

\end{document}